\documentclass[letterpaper,twocolumn,pre,aps,showpacs,groupedaddress,floatfix]{revtex4-1}
\usepackage[latin1]{inputenc}
\usepackage{bm}
\usepackage[usenames]{color}
\usepackage{multirow}
\usepackage{amssymb}
\usepackage{amsbsy}
\usepackage{amsmath}
\usepackage{stmaryrd}
\usepackage{graphicx}
\usepackage{epsfig}
\usepackage{placeins}
\usepackage{ulem}
\usepackage{braket}
\usepackage[colorlinks,linkcolor=blue,citecolor=blue,urlcolor=blue]{hyperref}

\newcommand{\sscript}{\text{neq}}

\makeatletter

\begin{document}

\title{Relation between far-from-equilibrium dynamics and equilibrium 
correlation functions for binary operators}

\author{Jonas Richter}
\email{jonasrichter@uos.de}

\author{Robin Steinigeweg}
\email{rsteinig@uos.de}
\affiliation{Department of Physics, University of Osnabr\"uck, D-49069 Osnabr\"uck, Germany}

\date{\today}



\begin{abstract}

Linear response theory (LRT) is one of the main approaches to the dynamics of
quantum many-body systems. However, this approach has limitations and requires,
e.g., that the initial state is (i) mixed and (ii) close to equilibrium.
In this paper, we discuss these limitations and study the 
nonequilibrium dynamics for a certain class of properly prepared initial 
states. Specifically, we consider thermal states of the quantum system in 
the presence of an additional static force which, however, become 
nonequilibrium states when this static force is eventually removed. While for 
weak forces the relaxation dynamics is well captured by LRT, much less 
is known in the case of strong forces, i.e., initial states far away 
from equilibrium. Summarizing our main results, we unveil that, for high 
temperatures, the nonequilibrium dynamics of so-called binary operators is 
always generated by an equilibrium correlation function. In particular, 
this statement holds true for states in the far-from-equilibrium limit, 
i.e., outside the linear response regime. In addition, we confirm our  
analytical results by numerically studying the dynamics of local fermionic 
occupation numbers and local energy densities in the spin-$1/2$ Heisenberg 
chain. Remarkably, these simulations also provide evidence that our results 
qualitatively apply in a more general setting, e.g., in the anisotropic XXZ 
model where the local energy is a non-binary operator, as well as for a wider 
range of temperature. Furthermore, exploiting the concept of quantum 
typicality, all of our findings are not restricted to mixed states, 
but are valid for pure initial states as well.

\end{abstract}

\maketitle


\section{Introduction}
Statistical physics provides a universal concept for the
calculation of equilibrium properties of many-body quantum systems. This
surprisingly simple and remarkably successful concept is to properly choose one
of the textbook statistical ensembles. Furthermore, various analytical and
numerical methods are available to carry out the actual calculation for a
specific physical model, see e.g.\ \cite{johnston2000, schollwoeck2005,
prelovsek2013}. Out of equilibrium, however, such a universal concept is
absent. This absence is not least related to the diversity of nonequilibrium
situations. On the one hand, there can be driving by time-dependent protocols
\cite{eckardt2005, lazarides2014, campisi2011} and by heat baths or particle
reservoirs at unequal temperatures or chemical potentials \cite{mejiamonasterio2005,
michel2008, znidaric2011}. In strictly isolated situations, on the other hand,
a variety of initial states can be prepared. These initial states can be mixed
or pure, entangled or non-entangled, and close to or far away from equilibrium.

Quantum many-body systems in strict isolation have experienced an upsurge of
interest in recent years, also due to the advent of cold atomic gases
\cite{langen2015}, the discovery of many-body localized phases
\cite{nandkishore2015}, and the invention of powerful numerical techniques such
as density matrix renormalization group \cite{schollwoeck2005}. In particular,
understanding the existence of equilibration and thermalization has seen
substantial progress \cite{eisert2015, dalessio2016} by as fascinating concepts
as eigenstate thermalization \cite{deutsch1991, srednicki1994, rigol2008} and
typicality of pure states \cite{gemmer2003, goldstein2006, popescu2006,
reimann2007, bartsch2009, bartsch2011, sugiura2012, sugiura2013, elsayed2013,
hams2000, iitaka2003, iitaka2004, white2009, monnai2014}. However, much less is
known on the route to equilibrium as such \cite{reimann2016, garciapintos2017}.
A widely used approach to the full time-dependent relaxation process is linear
response theory (LRT) \cite{kubo1991}. While this highly developed theory
predicts the dynamics of expectation values on the basis of correlation
functions, the calculation of these correlation functions can be a challenge in
practice, see e.g.\ \cite{prosen2011, prosen2013, karrasch2012, steinigeweg2014,
steinigeweg2015}. In addition to this practical issue, LRT as such has
limitations and requires, e.g., that the initial state is (i) mixed and
(ii) close to equilibrium.

In this situation, our paper takes a fresh perspective and studies the 
nonequilibrium dynamics for a certain class of initial states. To be precise, 
we consider thermal states of the quantum system in the presence of an 
additional static force. However, when this static force is eventually removed, 
these states become nonequilibrium states of the remaining Hamiltonian. 
Moreover, depending on the strength of the external force, they can be prepared 
close to as well as far away from equilibrium at arbitrary temperature. 
On the one hand, in the case of a weak force, the resulting dynamics is well 
captured by LRT. On the other hand, much less is known for strong 
forces, i.e., initial states far away from equilibrium. While the 
preparation of the initial states in principle does not require a specific type 
of observable, we here focus on so-called binary operators. For such operators 
and high temperatures, we unveil that the nonequilibrium dynamics is 
always generated by a single correlation function evaluated exactly at 
equilibrium. In particular, this statement holds true for states in the 
far-from-equilibrium limit, i.e., outside the linear response regime. In 
addition, we confirm our analytical results by numerically studying the 
dynamics of local fermionic occupation numbers \cite{fabricius1998, 
steinigeweg2017_1, steinigeweg2017_2, karrasch2017} and local energy densities 
\cite{karrasch2017, karrasch2014} in the spin-$1/2$ Heisenberg chain. 
Remarkably, these simulations also provide evidence that our results 
qualitatively apply in a more general setting, e.g., in the anisotropic XXZ 
model where the local energy is a non-binary operator, as well as for a wider 
range of temperature. Furthermore, exploiting the concept of quantum 
typicality, all of our findings are not restricted to mixed states, 
but are valid for pure initial states as well.

This paper is structured as follows: In Sec.\ \ref{Sec::StaticF} we introduce our nonequilibrium setup. 
We continue to discuss analytical results for this setup in Sec.\ \ref{Sec::NED}. 
In Sec.\ \ref{Sec::DQT} we present the numerical approach which is employed to 
illustrate our findings in Sec.\ \ref{Sec:NI}. We summarize and conclude in Sec.\ \ref{Sec:Conc}.


\section{Response to a Static Force}\label{Sec::StaticF}
We start by considering a quantum system
described by a Hamiltonian $\cal H $ which is in contact with a (weakly coupled
and macroscopically large) heat bath at temperature $T = 1/\beta$. Furthermore,
this quantum system is affected by a static force which gives rise to an
additional potential energy described by an operator ${\cal O}_l$ \cite{force,  
luttinger1964, zwanzig1965, gemmer2006}. (The subscript $l$ indicates 
that we have local operators in mind. Later there will be also other
operators ${\cal O}_{l'}$.) For such a situation, thermalization to the density
matrix
\begin{equation} \label{eq:rho_neq}
\rho_\sscript = \frac{e^{-\beta({\cal H} - \varepsilon {\cal O}_l)}}{{\cal
Z}_\sscript}
\end{equation}
emerges, where ${\cal Z}_\sscript = \text{Tr} [ e^{-\beta ({\cal H} -
\varepsilon {\cal O}_l)} ]$ is the partition function and the parameter
$\varepsilon$ denotes the strength of the static force. Eventually, this force
and the heat bath are both removed. This setup might be seen as a type of
quantum quench as well \cite{karrasch2014, Essler2014}. Then, $\rho_\sscript$ 
in Eq.\ \eqref{eq:rho_neq} is no equilibrium state of the remaining Hamiltonian 
$\cal H$ such that it evolves in time according to the von-Neumann 
equation for this Hamiltonian,
\begin{equation}
\rho_\sscript(t) = 
e^{-i\mathcal{H}t} \, \rho_\sscript \, e^{i\mathcal{H}t} \, .
\end{equation}

If $\varepsilon$ is a small parameter, the exponential in Eq.\  
\eqref{eq:rho_neq} can be expanded according to \cite{kubo1991, bartsch2017}
\begin{equation} \label{eq:rho_expansion}
\rho_\sscript = \rho_\text{eq} \Big ( 1 + \varepsilon \int_0^\beta \text{d}
\beta'\ e^{\beta' {\cal H}}\ \Delta {\cal O}_l\ e^{-\beta' {\cal H }}  +
\varepsilon^2 \ldots \Big )\ ,
\end{equation}
where $\Delta {\cal O}_l = {\cal O}_l - \langle {\cal O}_l \rangle_\text{eq}$
and $\langle \bullet \rangle_\text{eq} = \text{Tr} [ \rho_\text{eq} \bullet ]$ 
denotes the {\it equilibrium} expectation value with
\begin{equation}
\rho_\text{eq} = \frac{e^{-\beta {\cal H}}}{\mathcal{Z}_\text{eq}}\ , 
\end{equation}
and ${\cal Z}_\text{eq} = \text{Tr} [e^{-\beta {\cal H}} ]$. Hence, for small 
values of $\varepsilon$, the dynamical expectation value $\langle
{\cal O}_{l'}(t) \rangle_\sscript = \text{Tr} [ \rho_\sscript 
(t) \mathcal{O}_{l'} ]$ of some (other) operator 
$\mathcal{O}_{l'}$ can be written as
\begin{equation} \label{eq:LRT}
\langle {\cal O}_{l'}(t) \rangle_\sscript = \langle {\cal O}_{l'}
\rangle_\text{eq} + \varepsilon \, \chi_{l,l'}(t) \ ,
\end{equation}
where the susceptibility (or relaxation function) 
$\chi_{l,l'}(t)$ is given by \cite{kubo1991}, 
\begin{equation}\label{Eq::Chi}
\chi_{l,l'}(t) = \int_0^\beta d\beta'\ \text{Tr}[\rho_\text{eq} \,
\Delta {\cal O}_l(-i\beta') \, {\cal O}_{l'}(t)]\ . 
\end{equation}
Note that the expansion in Eq.\ \eqref{eq:rho_expansion} is known
to converge because all expressions are analytical and the operators involved
have bounded spectra. Equation \eqref{eq:LRT} reflects a central statement of 
LRT, i.e., for  small values of $\varepsilon$ the response is linear in 
$\varepsilon$. However, when $\varepsilon$ is increased to large values, 
higher-order terms are expected to become non-negligible \cite{peterson1967}.

\begin{figure}[tb]
\centering
\includegraphics[width=0.9\columnwidth]{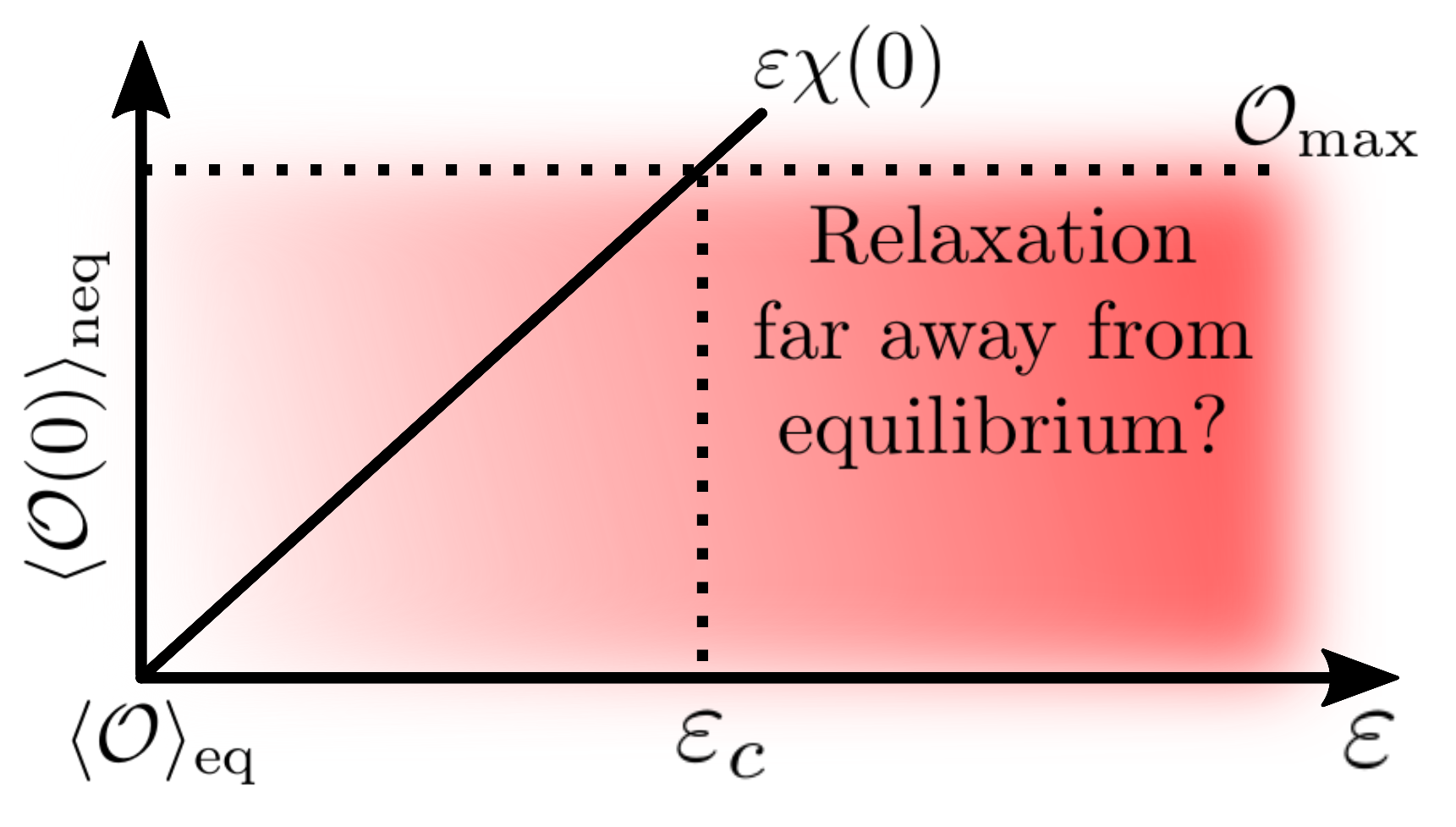}
\caption{(Color online) While for weak external forces, the initial 
expectation value $\langle {\cal O}(0)\rangle_\text{neq}$ follows 
the linear prediction in Eq.\ \eqref{eq:LRT}, this prediction has to break down 
at the latest for a perturbation of strength $\varepsilon_\text{c}$, 
cf.\ Eq.\ \eqref{Eq::Ec}. Here, we study the relaxation dynamics of initial  
states in the entire regime close to as well as far away from equilibrium.}
\label{Fig0}
\end{figure}

By tuning the strength of the external force, it is possible to  
prepare initial states which can be close to as well as far away from 
equilibrium. On the one hand, in the limit $\varepsilon \to 0$, one
naturally finds $\rho_\sscript \rightarrow \rho_\text{eq}$. On the other hand,
in the limit $\varepsilon \to \infty$, the density matrix $\rho_\sscript$ in 
Eq.\ \eqref{eq:rho_neq} acts as a projector on the eigenstates of ${\cal O}_l$ 
with the largest eigenvalue ${\cal O}_{l, \text{max}}$. Thus, the 
initial expectation value $\langle {\cal O}_{l'=l}(0)
\rangle_\sscript$ reads
\begin{equation} \label{eq:O_max}
\lim_{\varepsilon \to \infty} \langle {\cal O}_{l'=l}(0) \rangle_\sscript =
{\cal O}_{l,\text{max}}\ .
\end{equation}
In particular, comparing Eqs.\ \eqref{eq:LRT} and \eqref{eq:O_max} suggests 
that LRT has to break down at the latest for a perturbation of 
strength 
\begin{equation}\label{Eq::Ec}
\varepsilon_\text{c} = \frac{{\cal O}_{l,\text{max}} - \langle {\cal O}_{l'=l} 
\rangle_\text{eq}}{\chi_{l,l'=l}}\ .
\end{equation}
A central aspect of this work is to study the relaxation of  
expectation values $\langle {\cal O}_{l'}(t) \rangle_\sscript$ 
outside the regime of linear response, i.e., beyond the validity range of Eq.\ 
\eqref{eq:LRT}, see Fig.\ \ref{Fig0}.

\section{Nonequilibrium dynamics and correlation functions}\label{Sec::NED}

\subsection*{High temperatures and arbitrary perturbations}

Although the states $\rho_\sscript$ in Eq.\ \eqref{eq:rho_neq} do not 
require a specific type of operator, we here restrict ourselves to 
so-called binary operators ${\cal O}_l = c_1 ({\cal P} + c_2)$, where all 
eigenvalues of ${\cal P}$ are either $0$ or $1$, and $c_1$ and $c_2$ are some 
real constants. Then, ${\cal P}^2 = {\cal P}$ and ${\cal P}$ describes a 
projection. Binary operators are, e.g., local fermionic occupation numbers 
\cite{fabricius1998, steinigeweg2017_1, steinigeweg2017_2, karrasch2017} and 
local energy densities \cite{karrasch2017, karrasch2014} in the isotropic 
Heisenberg spin-$1/2$ chain. (These two examples will be considered in our 
numerical simulations below.)

Moreover, let us focus on the regime of high temperatures and strong 
external forces, i.e., we want to consider the regime $\beta \to 0$ but $\beta 
\varepsilon > 0$. In this regime, we can write to good approximation 
\begin{equation}
\rho_\sscript \approx \frac{e^{\beta \varepsilon c_1 ({\cal 
P} + c_2)}}{\text{Tr}[e^{\beta \varepsilon c_1 ({\cal P} + c_2)}]} = 
\frac{e^{\beta \varepsilon c_1 {\cal P}}}{\text{Tr}[e^{\beta \varepsilon 
c_1 {\cal P}}]}\ . 
\end{equation}
From now on, to simplify notation, we absorb $c_1$ in the definition of 
$\varepsilon$. A Taylor expansion of this exponential, combined with the 
projection property ${\cal P}^i = {\cal P}$, then yields
\begin{align}
\rho_\sscript\ &\propto 1 + \beta \varepsilon {\cal P} + 
\tfrac{1}{2} \beta^2 \varepsilon^2 {\cal P}^2 + \tfrac{1}{6} \beta^3 
\varepsilon^3 {\cal P}^3 + \dots \nonumber \\
&= 1 + \beta \varepsilon {\cal P} + \tfrac{1}{2} \beta^2 \varepsilon^2 {\cal P} 
+ \tfrac{1}{6} \beta^3 \varepsilon^3 {\cal P} + \dots \nonumber \\
&= 1 + (e^{\beta \varepsilon} - 1) \, {\cal P}\ .\label{Eq::PP}
\end{align}
Next, let us focus the situation where one measures the relaxation of 
exactly the same observable which is used to prepare the initial state, 
i.e., we have ${\cal O}_{l'} = {\cal O}_l = c_1({\cal P} + c_2)$. Then, 
it follows from  Eq.\ \eqref{Eq::PP} that the time-dependent expectation value 
$\langle {\cal O}_{l'}(t) \rangle_\sscript = c_1 (\langle {\cal 
P}(t)\rangle_\sscript + c_2)$ is given by the relation
\begin{align} \label{eq:all_epsilon}
\langle {\cal P}(t)\rangle_\sscript = \frac{\langle {\cal P} 
\rangle_\text{eq} + (e^{\beta \varepsilon} - 1)\cdot\langle{\cal P} {\cal P}(t)
\rangle_\text{eq}}{1 + (e^{\beta \varepsilon} - 1)\cdot \langle {\cal P} 
\rangle_\text{eq}}\ .
\end{align}
Thus, at high temperatures, the nonequilibrium dynamics of binary 
operators is generated by the equilibrium correlation function $\langle {\cal 
P} {\cal P}(t) \rangle_\text{eq}$ for {\it all} $\varepsilon$, i.e., even 
beyond the regime of small perturbations. This prediction is a central result 
of our paper.

In particular, for small  $\varepsilon$, Eq.\ \eqref{eq:all_epsilon} 
can be linearized (Taylor expansion up to linear order around 
$\varepsilon = 0$) and becomes 
$\langle {\cal P}(t)\rangle_\sscript = \langle {\cal P} 
\rangle_\text{eq} + \varepsilon \, \chi_{{\cal P},{\cal P}}(t)$ with
\begin{equation}
\chi_{{\cal P},{\cal P}}(t) \approx \beta \Big( \langle {\cal P} {\cal 
P}(t) \rangle_\text{eq} - \langle {\cal P} \rangle_\text{eq}^2 \Big )\ ,
\end{equation}
as expected from LRT. 
It is worth pointing out that such a dynamical independence of
$\varepsilon$ can hardly be expected at low temperatures. There, the time 
dependence of $\chi_{{\cal P},{\cal P}}(t)$ is not just given by 
$\langle {\cal P} {\cal P}(t) \rangle_\text{eq}$, cf.\ Eq.\ 
\eqref{Eq::Chi}. 

\subsection*{Arbitrary temperatures and strong perturbations}

In addition to the above considerations for $\beta \to 0$, it is also 
instructive to study the limit of infinitely strong perturbations $\varepsilon 
\to \infty$ at arbitrary temperatures $\beta \geq 0$. In the following, we will 
again consider binary operators which fulfill ${\cal P}^2 = {\cal P}$.
As discussed in the context of Eq.\ \eqref{eq:O_max}, $\rho_\sscript$ in Eq.\
\eqref{eq:rho_neq} acts as projector in the limit $\varepsilon \to
\infty$. Hence, we find
\begin{equation}
\lim_{\varepsilon \to \infty} \rho_\sscript \propto {\cal P} \, e^{-\beta {\cal 
H}} \, {\cal P}
\end{equation}
 and 
\begin{equation}\label{Eq::Neu}
\lim_{\varepsilon \to \infty} \langle {\cal P}(t) \rangle_\text{neq} = 
\tilde{C}(t) = \frac{\langle {\cal P}{\cal P}(t){\cal 
P}\rangle_\text{eq}}{\langle {\cal P} \rangle_\text{eq}}\ ,
\end{equation}
which is valid for any temperature $\beta \geq 0$. In some cases, 
$\tilde{C}(t)$ can be connected to the usual correlation function $\langle{\cal 
P} {\cal P}(t) \rangle_\text{eq}$ in Eq.\ \eqref{eq:all_epsilon}. Specifically, 
one can require some sort of ``particle-hole symmetry'', i.e., invariance of 
$\tilde{C}(t)$ under
\begin{equation}
{\cal P}(t) \to 1 - {\cal P}(t) \, .
\end{equation}
(In fact, this requirement is fulfilled by the local fermionic occupation 
numbers in the XXZ spin-$1/2$ chain discussed later.) Exploiting this property, 
one can write
\begin{equation}
\tilde{C}(t) = \frac{\langle [1-{\cal P}][1-{\cal P}(t)][1-{\cal 
P}]\rangle_\text{eq}}{\langle {\cal P} \rangle_\text{eq}}\ .
\end{equation}
Multiplying out the brackets on the r.h.s.\ of this relation, using $\langle
{\cal P} \rangle_\text{eq} = 1/2$, and rearranging a bit yields
\begin{equation}
2 \langle {\cal P} \, {\cal P}(t) \, {\cal P} \rangle_\text{eq} = \langle {\cal 
P} \, {\cal P}(t)
\rangle_\text{eq} + \langle {\cal P}(t) \, {\cal P} \rangle_\text{eq}\ , 
\end{equation}
and, since $\langle {\cal P}(t) \, {\cal P} \rangle_\text{eq} = \langle {\cal 
P} \,
{\cal P}(t) \rangle_\text{eq}^*$,
\begin{equation}
\langle {\cal P} \, {\cal P}(t) \, {\cal P} \rangle_\text{eq} = \text{Re} \, 
\langle {\cal P} \, {\cal P}(t) \rangle_\text{eq}\ .
\end{equation}
Thus, as a consequence of this identity, we can rewrite the correlation  
function $\tilde{C}(t)$ as
\begin{equation}\label{eq:final}
\tilde{C}(t) = \text{Re} \, \frac{\langle {\cal P} \, {\cal P}(t)
\rangle_\text{eq}}{\langle {\cal P} \rangle_\text{eq}}\ .
\end{equation}
Comparing Eqs.\ \eqref{Eq::Neu} and \eqref{eq:final}, the 
nonequilibrium expectation value $\langle {\cal P}(t) \rangle_\text{neq}$
in the limit $\varepsilon \to \infty$ eventually recovers the real part of the  
equilibrium correlation function $\langle {\cal P} \, {\cal P}(t) 
\rangle_\text{eq}$ at {\it any} temperature. This is another main result.

\section{Dynamical typicality and pure-state propagation}\label{Sec::DQT}

Time-dependent expectation values of the form $\langle {\cal O}_{l'} (t) 
\rangle_\sscript = \text{Tr} [ \rho_\sscript(t) {\cal O}_{l'} ]$ 
can be calculated exactly, if the eigenstates and eigenvalues of the 
Hamiltonians ${\cal H} - \varepsilon {\cal O}_l$ and ${\cal H}$ are obtained 
from the exact diagonalization (ED) of finite systems. But, in addition to the 
main limitation set by the exponential growth of many-body Hilbert spaces, this 
procedure is also costly since it requires to perform the exact diagonalization 
of two operators. Therefore, we here proceed differently and rely on 
the concept of dynamical typicality (QT) \cite{gemmer2003, goldstein2006,
popescu2006, reimann2007, bartsch2009, bartsch2011, sugiura2012, sugiura2013,
elsayed2013, hams2000, iitaka2003, iitaka2004, white2009, monnai2014}. This 
concept states that a {\it single} pure state can have the same properties as
the ensemble density matrix. Precisely, the main idea is to replace the trace
$\text{Tr} [ \rho_\sscript \mathcal{O}_{l'}(t) ]$ by the 
scalar product $\bra{\phi} \rho_\sscript \mathcal{O}_{l'}(t)
\ket{\phi}$, where the pure state $\ket{\phi}$ is drawn at random according to 
the unitary invariant Haar measure \cite{bartsch2009, bartsch2011}. By the use 
of this replacement, the expectation value $\langle {\cal O}_{l'}(t) 
\rangle_\sscript$ can be written as 
\begin{equation} \label{eq:QT1}
\langle {\cal O}_{l'}(t) \rangle_\sscript = \bra{\psi_\sscript 
(t)} {\cal O}_{l'} \ket{\psi_\sscript(t)} + f(\ket{\phi})
\end{equation}
with the nonequilibrium pure state $\ket{\psi_\sscript(t)} = e^{-i 
\mathcal{H}t}\ket{\psi_\sscript(0)}$ and 
\begin{equation} \label{eq:state1}
\ket{\psi_\sscript(0)} = \frac{\sqrt{\rho_\sscript} \ket{\phi}}{\sqrt{\langle \phi
|\rho_\sscript|\phi \rangle}}\ .
\end{equation}
Note that the statistical error in Eq.\ \eqref{eq:QT1} scales as $f(\ket{\phi})  
\propto 1/d_\text{eff}^{1/2}$, where $d_\text{eff} = {\cal 
Z}_\sscript/e^{-\beta  E_\sscript}$ is the effective dimension of the Hilbert 
space and $E_\sscript$ denotes the energy of the ground state. Thus, as the 
size of a many-body quantum system is increased, $f(\ket{\phi})$ vanishes 
exponentially fast and can be neglected for medium system sizes already 
\cite{steinigeweg2014, steinigeweg2015} (cf.\ Appendix \ref{Sec:EA}  
and \ref{Sec:FSE}). For a recent discussion of dynamical typicality 
and similar classes of pure states, see Ref.\ \cite{reimann2018}. 
Note further, that the concept of typicality has also been recently 
used to study linear and nonlinear responses in a different but related setup 
\cite{Endo2018}.

Relying on the QT relation in Eq.\ \eqref{eq:QT1} we do not need to deal 
with density matrices and can consider pure states instead. Moreover, 
given the class of pure states \eqref{eq:state1} in combination with Eqs.\ 
\eqref{eq:all_epsilon} and \eqref{eq:final}, only a {\it single} state is 
required, compared to earlier calculations of correlation functions based on 
{\it two} (auxiliary) pure states \cite{steinigeweg2014,steinigeweg2015}.
While a forward propagation w.r.t.\ ${\cal H} - \varepsilon {\cal O}_l$ in 
imaginary time $\beta$ allows us to prepare $\ket{\psi_\sscript(0)}$,
another forward propagation w.r.t.\ ${\cal H}$ in real time $t$ allows us to
calculate $\ket{\psi_\sscript(t)}$. The main advantage of the  
pure-state approach comes from the fact that these propagations can be done by
iteratively solving the Schr\"odinger equation (in real and imaginary time),  
e.g., by a fourth-order Runge-Kutta scheme with a small time step $\delta t$
\cite{elsayed2013, steinigeweg2014, steinigeweg2015, herbrych2016}. This scheme
does not require exact diagonalization and, due to the fact that few-body
operators are relatively sparse, also the matrix-vector multiplications can be
implemented in a very memory-efficient way. Hence, in comparison  to exact
diagonalization, we can treat systems with much larger Hilbert spaces. Note that
also more sophisticated schemes can be applied such as Trotter decompositions or
Chebyshev polynomials \cite{steinigeweg2017_1, steinigeweg2017_2, weisse2006}.
However, for the purposes of our paper, i.e., the numerical illustration of our 
analytical results, Runge-Kutta will be sufficient.


\section{Numerical Illustration}\label{Sec:NI}

\begin{figure}[tb]
\centering
\includegraphics[width=0.95\columnwidth]{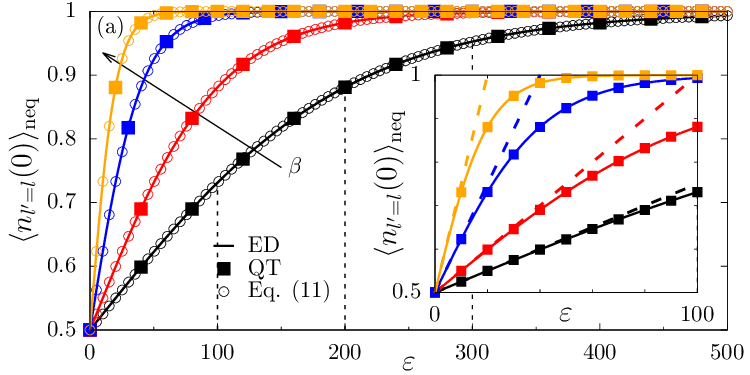}
\caption{(Color online) (a) Static expectation value $\langle n_{l'=l}(0)
\rangle_\sscript$ versus perturbation $\varepsilon$ for high temperatures
$\beta = 0.01$, $0.02$, $0.05$, and $0.1$ (arrow). Exact diagonalization (ED)
is compared to quantum typicality (QT). The analytical prediction in Eq.\
\eqref{eq:all_epsilon} is depicted for comparison. The vertical dashed  
lines indicate those values of $\varepsilon$ which will be used to study the 
nonequilibrium dynamics in the following. (b) Same data as in (a) but
shown for $\varepsilon \leq 100$  only. The LRT prediction in Eq.\ \eqref{eq:LRT}
is now indicated (dashed line). Parameters: $\Delta = 1$ and $\delta
t = 0.01$ as well as $L = 8$ (ED) and $L=24$ (QT).}
\label{fig:static}
\end{figure}


\subsection{Model} 
Next, we turn to our numerical simulations and
study, as an example, nonequilibrium dynamics in the XXZ spin-1/2 chain. 
The Hamiltonian of this chain reads (with periodic boundary conditions)
$\mathcal{H} = \sum_{l = 1}^L h_l$,
\begin{equation} \label{eq:H}
h_l = J \left(S_l^x S_{l+1}^x + S_l^y S_{l+1}^y + \Delta S_l^z S_{l+1}^z
\right)\ , 
\end{equation}
where $S_l^{x,y,z}$ are spin-$1/2$ operators at site $l$, $L$ is the number of
sites, $J = 1$ is the antiferromagnetic exchange coupling constant, and $\Delta$
is the anisotropy. By the use of the Jordan-Wigner transformation, this
Hamiltonian can be also mapped onto a one-dimensional model of spinless fermions
with interactions between nearest neighbors. In this picture, the operator $n_l
= S_l^z + 1/2$ becomes a local fermionic occupation number. Because such an
operator has only the two eigenvalues $0$ and $1$, it naturally  
fulfills the projection property $n_l^2 = n_l$. Consequently, we prepare the 
initial state $\rho_\sscript$ by using the choice ${\cal O}_l = n_l$ and 
subsequently measure the observable ${\cal O}_{l'} = n_{l'}$. In fact, we here 
restrict ourselves for simplicity to the single case $l = l'$. Note 
that our numerical simulations are performed for all subsectors of fixed 
magnetization $S^z = \sum_l S_l^z$ (or, in the fermionic language, all 
subsectors of fixed particle number $N = \sum_l n_l$).


\subsection{Results}

\subsection*{Static expectation values}

We now discuss our numerical results. First, we study the influence of the
perturbation $\varepsilon$ on static expectation values, i.e., we investigate
the validity range of the LRT prediction in Eq.\ \eqref{eq:LRT} at $t = 0$. In 
Fig.\ \ref{fig:static} (a) we show $\langle n_{l'=l}(0)
\rangle_\sscript$ for a wide range $\varepsilon \leq 500$, high temperatures 
$\beta = 0.01$, $0.02$, $0.05$, and $0.1$, as well as anisotropy $\Delta = 1$. 
At $\varepsilon = 0$, we have $\langle n_{l'=l}(0) \rangle_\sscript = 
\langle n_{l'=l} \rangle_\text{eq} = 1/2$. As $\varepsilon$ increases, we 
observe a linear growth of $\langle n_{l'=l}(0) \rangle_\sscript$ 
with $\varepsilon$. As depicted in Fig.\ \ref{fig:static} (b), this linear 
growth is very well described by the LRT prediction in Eq.\ \eqref{eq:LRT} and 
$\chi_{l,l'=l} = \beta/4$. For large $\varepsilon$, $\langle 
n_{l'=l}(0) \rangle_\sscript$ eventually saturates at the constant 
value $\langle n_{l'=l}(0) \rangle_\sscript = 1$. This saturation
is expected due to Eq.\ \eqref{eq:O_max} and the maximum eigenvalue $n_{l,
\text{max}} = 1$. In Figs.\ \ref{fig:static} (a) and (b), the QT relation 
in Eq.\ \eqref{eq:QT1} is additionally confirmed by a direct comparison with 
data from exact diagonalization. For all $\beta$ considered here, the  
statistical error $f(\ket{\phi})$ turns out to be 
negligibly small. Overall, the numerical results in Figs.\ \ref{fig:static} (a) 
and (b) confirm our analytical predictions. In particular, static LRT breaks 
down for sufficiently large $\varepsilon$. As an additional 
orientation, the vertical dashed lines in Fig.\ \ref{fig:static} (a) indicate 
those values of $\varepsilon$ which will be used to study the nonequilibrium 
dynamics in the following. Note that these values are chosen in such a way that 
we cover the whole range from states inside the linear regime, as well as 
initial states which are almost maximally perturbed.

\begin{figure}[tb]
\centering
\includegraphics[width=0.95\columnwidth]{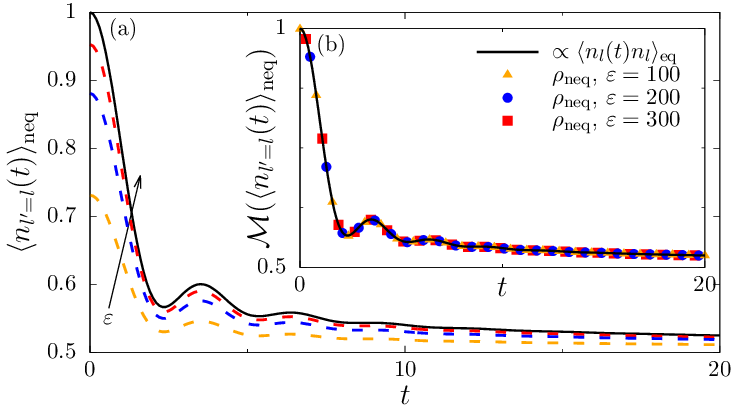}
\caption{(Color online) (a) Time evolution of the expectation value $\langle
n_{l'=l}(t) \rangle_\sscript$ for perturbations $\varepsilon = 100$, $200$,
and $300$ (arrow) at a high temperature $\beta = 0.01$. As a  
comparison, the (normalized) equilibrium correlation function $\langle n_l(t) 
n_l \rangle_\text{eq}$ is shown. (b) Data collapse of (a) using a simple 
linear map. Parameters: $L=28$ and $\delta t = 0.01$ as well as $\Delta = 1$.}
\label{fig:dynamic}
\end{figure}

\subsection*{Dynamical expectation values}

Next, we turn to dynamical expectation values, with the focus on a high
temperature $\beta = 0.01$. In Fig.\ \ref{fig:dynamic} (a) we depict
$\langle n_{l'=l}(t) \rangle_\sscript$ for perturbations $\varepsilon = 100$,
$200$, and $300$, as resulting from the QT relation \eqref{eq:QT1}. Further,  
as a comparison, we depict the (normalized) equilibrium correlation 
function $\langle n_{l}(t) n_l \rangle_\text{eq}$, which can be calculated 
by means of a pure-state approach as well \cite{steinigeweg2014, 
steinigeweg2015}. While we observe that all curves shown differ from each 
other, this observation is not surprising because, as illustrated in Fig.\
\ref{fig:static}, the initial values $\langle n_{l'=l}(0) \rangle_\sscript$
depend on $\varepsilon$. In view of this fact, we try a data collapse using the
simple map (see Appendix \ref{Sec:LM})
\begin{equation} \label{eq:rescaling}
{\cal M} (\langle n_{l'=l}(t) \rangle_\sscript) = a\ \langle n_{l'=l}(t)
\rangle_\sscript + b\ ,
\end{equation}
where the strictly {\it time-independent} coefficients $a$ and $b$ can be  
chosen as
\begin{equation}
 a = \frac{n_{l,max} - \langle n_{l'=l} \rangle_\text{eq}}{\langle n_{l'=l}
\rangle_\sscript - \langle n_{l'=l} \rangle_\text{eq}}\, , \quad b =  (1 - a) 
\langle n_{l'=l} \rangle_\text{eq}\ .
\end{equation}
Note that in the case of a traceless operator, $\langle \mathcal{O} 
\rangle_\text{eq} = 0$, this map essentially reduces to a normalization of the 
data to the initial value,
\begin{equation}
{\cal M} (\langle 
\mathcal{O}(t) \rangle_\sscript) \propto \frac{\langle \mathcal{O}(t) 
\rangle_\sscript}{\langle \mathcal{O}(0) \rangle_\sscript}\ .
\end{equation}
Due to our discussion in the context of Eqs.\ \eqref{eq:LRT} and  
\eqref{eq:all_epsilon}, such a linear map is reasonable for small and large  
$\varepsilon$. And indeed, as shown in Fig.\ \ref{fig:dynamic} (b), the 
rescaled curves lie on top of each other for all values of $\varepsilon$. This 
finding is a main result of our paper, as it clearly confirms our analytical 
prediction in Eq.\ \eqref{eq:all_epsilon} that the dynamical behavior at high 
temperatures, even outside the linear response regime, is generated by 
a single equilibrium correlation function, at least for binary operators. 
Moreover, it is worth noting that the data is presented for system 
size $L = 28$, substantially beyond what is possible in conventional ED 
\cite{fabricius1998}.

\begin{figure}[tb]
\centering
\includegraphics[width=0.95\columnwidth]{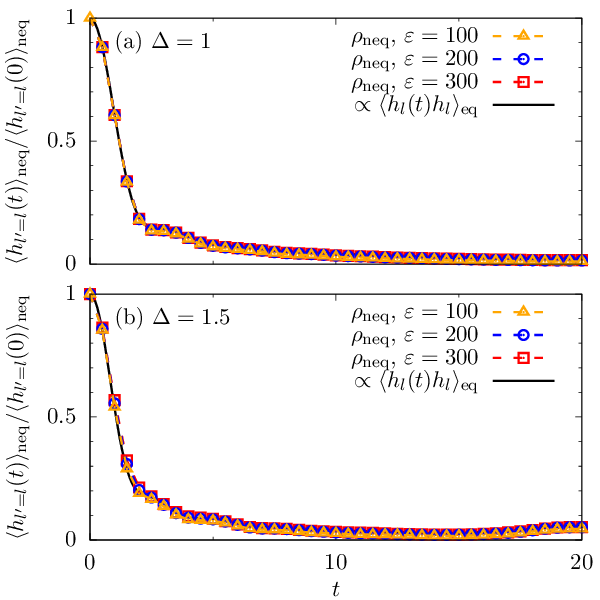}
\caption{(Color online) Time evolution of the expectation value $\langle
h_{l'=l}(t) \rangle_\sscript$ for perturbations $\varepsilon = 100$, $200$,
and $300$ (arrow) at a high temperature $\beta = 0.01$ for (a) $\Delta = 1$ 
and (b) $\Delta = 1.5$. As a comparison, the (normalized)
equilibrium correlation function $\langle h_l(t) h_l \rangle_\text{eq}$ is shown. 
Other parameters: $L=28$ ($\Delta = 1$), $L = 24$ ($\Delta = 1.5$), and $\delta 
t = 0.01$.}
\label{fig:dynamic_2}
\end{figure}

The local fermionic occupation numbers $n_l$ considered so far appear in various
physical models, not just in the Heisenberg model or in one dimension. To
demonstrate, however, that our results are not restricted to such $n_l$, we
extend our analysis to other operators and consider the local energy density
$h_l$ in Eq.\ \eqref{eq:H}. This local energy density is a spin dimer. For
anisotropy $\Delta = 1$, this dimer features a triplet state
($\ket{\uparrow \uparrow}$, $\ket{\downarrow \downarrow}$, $\ket{\uparrow
\downarrow} + \ket{\downarrow \uparrow}$) with energy $E_\text{t} =1/4$ and a
singlet ground state ($\ket{\uparrow \downarrow} - \ket{\downarrow \uparrow}$)
with energy $E_\text{s} = -3/4$. Hence, Eq.\ (\ref{eq:all_epsilon}) 
holds for the projector ${\cal P} = h_l-E_\text{s}$, and dynamical 
independence of the perturbation is obviously expected for the choice ${\cal 
O}_l = h_l$ as well. Indeed, as shown in Fig.\  \ref{fig:dynamic_2} (a), the 
rescaled expectation values $\langle h_{l'=l}(t) \rangle_\sscript$ for 
different perturbations $\varepsilon$ all lie on top of each other. Therefore, 
our results are clearly valid for a larger class of binary operators.

\subsection*{Nonbinary operators}

The natural question arises if and to what extend our results also 
apply to nonbinary operators. To work towards an answer of this question let 
us also consider a larger anisotropy $\Delta = 1.5$. In this case, the local 
energy density $h_l$ becomes a  nonbinary operator (since the 
degeneracy of the triplet state is partially lifted). As a 
consequence, ${\cal P} = h_l - E_\text{s}$ does not fulfill the projection 
property, ${\cal P}^2 \neq {\cal P}$, and our derivation from Eq.\ 
\eqref{Eq::PP} will certainly break down. Thus, we cannot expect {\it a priori} 
that the relaxation dynamics is still independent of the perturbation 
$\varepsilon$. However, as depicted in Fig.\ \ref{fig:dynamic_2} (b), the 
numerical results reveal that even for this example of a nonbinary operator, 
the dynamics are still very well generated by a single correlation function.
Although this finding cannot be explained within the framework discussed in 
the paper at hand, it suggests that there might exist a more general theory for 
arbitrary operators.

\subsection*{Lower temperatures}

\begin{figure}[tb]
\centering 
\includegraphics[width = 0.95\columnwidth]{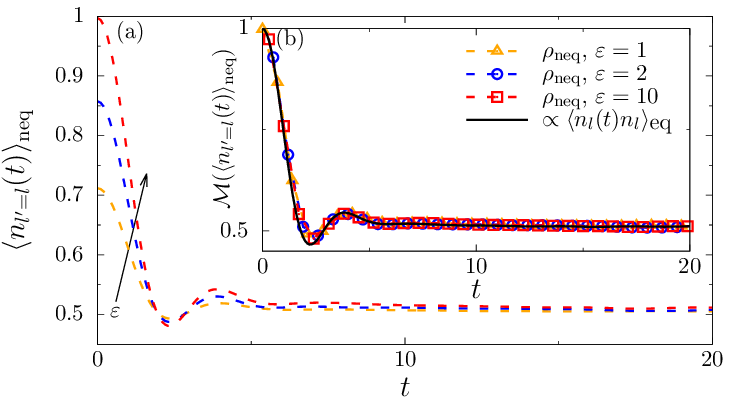}
\caption{(Color online) (a) Time evolution of the expectation value $\langle
n_{l'=l}(t) \rangle_\sscript$ for perturbations $\varepsilon = 1$, $2$, and
$10$ (arrow) at a lower temperature $\beta = 1$. (b) Rescaling of (a) using a
simple linear map. As a comparison, the (normalized) equilibrium 
correlation function $\langle n_l(t) n_l \rangle_\text{eq}$ is shown.
Parameters: $L = 24$, $\delta t = 0.01$, and $\Delta = 1$.}
\label{fig:lower_T}
\end{figure}

In the context of Eqs.\ \eqref{Eq::PP} and \eqref{eq:all_epsilon} we 
proved that at {\it high} temperatures the nonequilibrium dynamics of binary 
operators is already characterized by a single equilibrium correlation function. 
Moreover, we numerically confirmed this prediction in Figs.\ \ref{fig:dynamic} 
and \ref{fig:dynamic_2} (a) by demonstrating that the relaxation curves for 
different perturbation strengths $\varepsilon$ lie on top of each other after 
the simple linear map. Clearly, such an independence of $\varepsilon$ can 
hardly be expected at low temperatures. However, it may occur over a wider 
range of high temperatures $\beta > 0.01$. Thus, we redo the calculation for 
$\langle n_{l'=l}(t) \rangle_\sscript$ in Figs.\ \ref{fig:dynamic} (a) and (b) 
for the lower temperature $\beta = 1$ and depict the corresponding results in 
Figs.\ \ref{fig:lower_T} (a) and (b). For comparison, we also show the 
(normalized) correlation function $\langle n_{l}(t) n_{l} 
\rangle_\text{eq}$ at $\beta = 1$. While the data collapse is 
certainly not as good as before, it is still convincing.

\section{Conclusion}\label{Sec:Conc}

To summarize, we have studied the nonequilibrium dynamics for a class 
of initial states resulting from a certain type of quench. Specifically, we 
considered thermal states of the system in the presence of an additional 
external force which, however, become nonequilibrium states when this force 
is eventually removed. Moreover, by tuning the strength of the external force, 
these states can be prepared close to as well as far away from equilibrium, 
i.e., inside as well as outside the linear response regime, at arbitrary 
temperature.

In particular, we discussed the case of so-called binary operators
and specific examples for such observables. For these operators we  
proved that the nonequilibrium dynamics at high temperatures is characterized 
by a single correlation function evaluated exactly at equilibrium, even in the 
case of arbitrarily strong perturbations. This analytical result can also help 
to understand earlier numerical experiments where nonequilibrium dynamics has 
been related to linear response theory \cite{karrasch2017, Karrasch2013}.

In order to verify our results, we employed an efficient numerical  
approach based on the concept of dynamical typicality and studied the 
nonequilibrium dynamics in the spin-$1/2$ XXZ chain. In addition to confirming 
our analytical predictions, these simulations also provided evidence that 
our results might be (at least qualitatively) applicable in a much wider 
context, i.e., lower temperatures as well as more general types of operators. 
Albeit our numerical examples are certainly very different, it is in this 
context also worth mentioning two very recent experiments where universal 
dynamics in a far-from-equilibrium situation has been observed 
\cite{Prufer2018, Erne2018}.

Promising future directions of research include the analysis of non-binary
operators and low temperatures (in more detail), as well as the 
application of the pure-state approach to specific questions in many-body 
quantum systems. In this context, the study of nonequilibrium 
dynamics, transport properties, and thermalization in disordered systems is one 
important example \cite{Richter2018_2}. Moreover, a very recent work 
\cite{richter2018} shows that our results also hold true for a wider class of 
generic observables and models, as long as the off-diagonal eigenstate 
thermalization hypothesis applies.


\subsection*{Acknowledgments}
This work has been funded by the Deutsche Forschungsgemeinschaft (DFG) - STE 
2243/3-1. We sincerely thank J.\ Gemmer, P. Reimann, J. Herbrych, and 
the members of the DFG Research Unit FOR 2692 for fruitful discussions.


\appendix


\begin{figure}[tb]
\centering 
\includegraphics[width = 0.9\columnwidth]{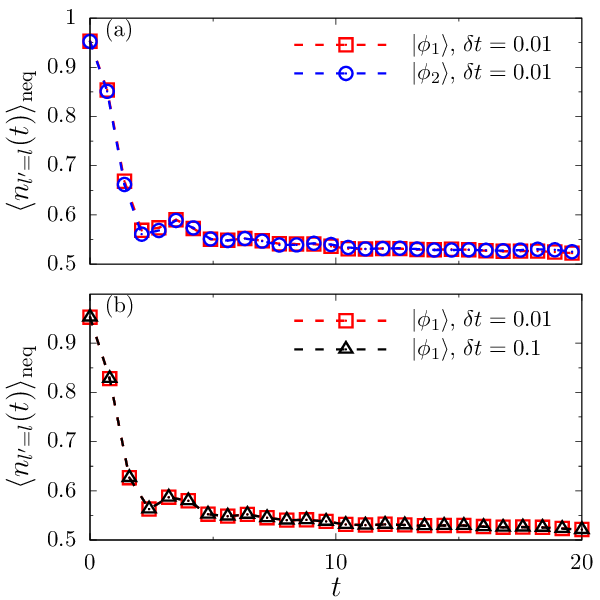}
\caption{(Color online) Analysis of errors.  Time evolution of the expectation
value $\langle n_{l'=l}(t)\rangle_\sscript$ for perturbation $\varepsilon =
300$, temperature $\beta = 0.01$, anisotropy $\Delta = 1$, and system size $L
= 16$, as resulting for: (a) two different reference states $\ket{\phi_1}$ and
$\ket{\phi_2}$, (b) two different Runge-Kutta time steps $\delta t = 0.01$ and
$0.1$.}
\label{fig:Error_Analysis}
\end{figure}

\section{Linear Map}\label{Sec:LM}

The time-independent coefficients of the linear map in Eq.\ \eqref{eq:rescaling}
follow from two conditions. The first condition is about the initial value,
i.e.,
\begin{equation}
a\ \langle n_{l'=l}(t = 0) \rangle_\sscript + b = c_0\ .
\end{equation}
The second condition is about the long-time value, i.e.,
\begin{equation}
a\ \langle n_{l'=l}(t \to \infty) \rangle_\sscript + b = c_\infty\ .
\end{equation}
We {\it choose} $c_0 = n_{l,max}$ and $c_\infty = \langle n_{l'=l}
\rangle_\text{eq}$ and get the parameters
\begin{equation}
a = \frac{n_{l,max} - \langle n_{l'=l} \rangle_\text{eq}}{\langle n_{l'=l}
\rangle_\sscript - \langle n_{l'=l} \rangle_\text{eq}}
\end{equation}
and $b =  (1 - a) \langle n_{l'=l} \rangle_\text{eq}$. Other choices are also
possible, of course.


\begin{figure}[b]
\centering 
\includegraphics[width = 0.9\columnwidth]{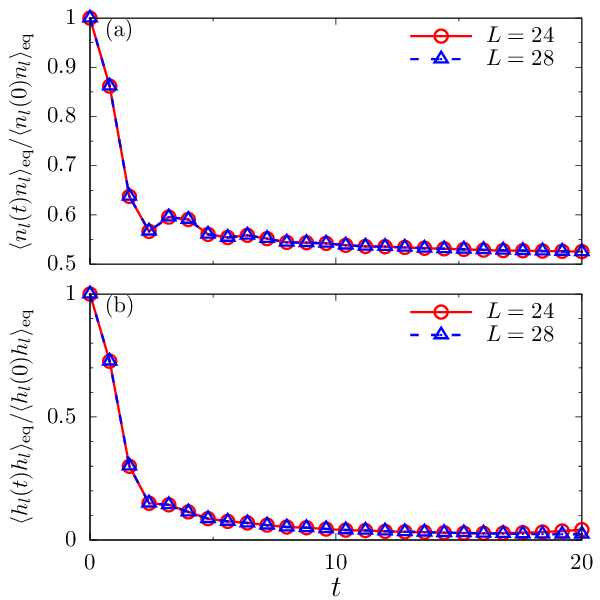}
\caption{(Color online) Time evolution of the (normalized) equilibrium 
correlation functions (a) $\langle n_{l}(t)n_l\rangle_\text{eq}$ and (b) 
$\langle h_{l}(t)h_l\rangle_\text{eq}$ for anisotropy $\Delta = 1$ and two 
different system sizes $L = 24$ and $L = 28$.}
\label{fig:finite_size}
\end{figure}

\section{Error Analysis}\label{Sec:EA}

Eventually, let us further comment on the accuracy of our pure-state approach,
i.e., of the typicality relation in Eq.\ \eqref{eq:QT1} of the main text. 
While the comparison with exact diagonalization in Fig.\
\ref{fig:static} has illustrated this accuracy already for static expectation
values, there is another convenient way to demonstrate the smallness of
statistical errors. This way is the comparison of results for two or even more
instances of the reference state $\ket{\phi}$ in Eq.\ \eqref{eq:state1}. 
Note that a single instance of this pure state is $\ket{ \phi} = \sum_k c_k 
\ket{\varphi_k}$, where the real and imaginary part of the complex coefficients 
$c_k$ are drawn at random according to a Gaussian distribution with zero mean 
and $\ket{\varphi_k}$ is the Ising basis.

In Fig.\ \ref{fig:Error_Analysis} (a) we exemplarily compare the dynamical
expectation values $\langle n_{l'=l}(t) \rangle_\sscript$ for two different
random realizations $\ket{\phi_1}$ and $\ket{\phi_2}$. For both realizations,
we use the same perturbation $\varepsilon = 300$, temperature $\beta = 0.01$,
and system size $L = 16$. Since the two curves coincide almost perfectly, we
can conclude that statistical errors are indeed very small, even for chains
with only $L = 16$ sites. Because these errors decrease exponentially fast as
the number of sites increases, our calculations for larger system sizes $L = 28$
in the main text can be considered as practically exact.

Eventually, we compare in Fig.\ \ref{fig:Error_Analysis} (b) the time evolution of
the expectation value $\langle n_{l'=l}(t) \rangle_\sscript$ for the same set
of parameters but two different Runge-Kutta time steps $\delta t = 0.01$ and
$0.1$. As the curves do not differ for these two choices, the time step $\delta
t = 0.01$ chosen throughout our paper is certainly small enough to ensure
negligibly small numerical errors.


\section{Finite-Size Effects}\label{Sec:FSE}

While it is evident from Fig.\ \ref{fig:static} that finite-size effects are
negligibly small for static expectation values, let us also comment briefly on
finite-size effects for dynamical expectation values. To this end, we compare
in Fig.\ \ref{fig:finite_size} numerical data for two different system sizes
$L=24$ and $L=28$. Apparently, for $\langle n_{l}(t) n_l \rangle_\text{eq}$ in
Fig.\ \ref{fig:finite_size} (a), both curves coincide with each other, at least
for all times $t \leq 20$ depicted. (Data from exact diagonalization can be
found for small $L=16 \ll 28$ in \cite{fabricius1998}). For $\langle h_{l}(t)h_l
\rangle_{\text{eq}}$ in Fig.\ \ref{fig:finite_size} (b), one can see minor
deviations at times $t \sim 20$.

However, we should stress that such finite-size effects do not affect the
conclusions in the main text. In fact, all relations discussed do not require
the convergence to the thermodynamic limit.



\begin{thebibliography}{99}

\bibitem{johnston2000}
D. C. Johnston, R. K. Kremer, M. Troyer, X. Wang, A. Kl\"umper, S. L. Bud'ko, A. F. Panchula, and P. C. Canfield,
Phys. Rev. B {\bf 61}, 9558 (2000).

\bibitem{schollwoeck2005} U. Schollw\"ock,
Rev. Mod. Phys. {\bf 77}, 259 (2005);
Ann. Phys. {\bf 326}, 96 (2011).

\bibitem{prelovsek2013}
P. Prelov\v{s}ek and J. Bon\v{c}a,
{\it Ground State and Finite Temperature Lanczos Methods},
Solid-State Sciences {\bf 176} (Springer, Berlin, 2013).

\bibitem{eckardt2005}
A. Eckardt, C. Weiss, and M. Holthaus,
Phys. Rev. Lett. {\bf 95}, 260404 (2005).

\bibitem{lazarides2014}
A. Lazarides, A. Das, and R. Moessner,
Phys. Rev. Lett. {\bf 112}, 150401 (2014).

\bibitem{campisi2011}
M. Campisi, P. H\"anggi, and P. Talkner,
Rev. Mod. Phys. {\bf 83}, 771 (2011).

\bibitem{mejiamonasterio2005}
C. Mej\'{i}a-Monasterio, T. Prosen, and G. Casati,
EPL (Europhys. Lett.) {\bf 72}, 520 (2005).

\bibitem{michel2008}
M. Michel, O. Hess, H. Wichterich, and J. Gemmer,
Phys. Rev. B {\bf 77}, 104303 (2008).

\bibitem{znidaric2011}
M. \v{Z}nidari\v{c},
Phys. Rev. Lett. {\bf 106}, 220601 (2011).

\bibitem{langen2015}
T. Langen, R. Geiger, and J. Schmiedmayer,
Annu. Rev. Condens. Matter Phys. {\bf 6}, 201 (2015).

\bibitem{nandkishore2015}
R. Nandkishore and D. A. Huse,
Annu. Rev. Condens. Matter Phys. {\bf 6}, 15 (2015).

\bibitem{eisert2015}
J. Eisert, M. Friesdorf, and C. Gogolin,
Nature Phys. {\bf 11}, 124 (2015).

\bibitem{dalessio2016}
L. D'Alessio, Y. Kafri, A. Polkovnikov, and M. Rigol,
Adv. Phys. \textbf{65}, 239 (2016).

\bibitem{deutsch1991}
J. M. Deutsch,
Phys. Rev. A {\bf 43}, 2046 (1991).

\bibitem{srednicki1994}
M. Srednicki,
Phys. Rev. E {\bf 50}, 888 (1994).

\bibitem{rigol2008}
M. Rigol, V. Dunjko, and M. Olshanii,
Nature {\bf 452}, 854 (2008).

\bibitem{gemmer2003}
J. Gemmer and G. Mahler,
Eur. Phys. J. B {\bf 31}, 249 (2003).

\bibitem{goldstein2006}
S. Goldstein, J. L. Lebowitz, R. Tumulka, and N. Zangh\`{\i},
Phys. Rev. Lett. {\bf 96}, 050403 (2006).

\bibitem{popescu2006}
S. Popescu, A. J. Short, and A. Winter,
Nature Phys. {\bf 2}, 754 (2006).

\bibitem{reimann2007}
P. Reimann,
Phys. Rev. Lett. {\bf 99}, 160404 (2007).

\bibitem{bartsch2009}
C. Bartsch and J. Gemmer,
Phys. Rev. Lett. {\bf 102}, 110403 (2009).

\bibitem{bartsch2011}
C. Bartsch and J. Gemmer,
EPL (Europhys. Lett.) {\bf 96}, 60008 (2011).

\bibitem{sugiura2012}
S. Sugiura and A. Shimizu,
Phys. Rev. Lett. {\bf 108}, 240401 (2012).

\bibitem{sugiura2013}
S. Sugiura and A. Shimizu,
Phys. Rev. Lett. {\bf 111}, 010401 (2013).

\bibitem{elsayed2013}
T. A. Elsayed and B. V. Fine,
Phys. Rev. Lett. {\bf 110}, 070404 (2013).

\bibitem{hams2000}
A. Hams and H. De Raedt,
Phys. Rev. E {\bf 62}, 4365 (2000).

\bibitem{iitaka2003}
T. Iitaka and T. Ebisuzaki,
Phys. Rev. Lett. {\bf 90}, 047203 (2003).

\bibitem{iitaka2004}
T. Iitaka and T. Ebisuzaki,
Phys. Rev. E {\bf 69}, 057701 (2004).

\bibitem{white2009}
S. R. White,
Phys. Rev. Lett. {\bf 102}, 190601 (2009).

\bibitem{monnai2014}
T. Monnai and A. Sugita,
J. Phys. Soc. Jpn. {\bf 83}, 094001 (2014).

\bibitem{reimann2016}
P. Reimann,
Nature Commun. {\bf 7}, 10821 (2016).

\bibitem{garciapintos2017}
L. P. Garc\'{\i}a-Pintos, N. Linden, A. S. L. Malabarba, A. J. Short, and A. Winter,
Phys. Rev. X {\bf 7}, 031027 (2017).

\bibitem{kubo1991}
R. Kubo, M. Toda and, N. Hashitsume,
{\it Statistical Physics II: Nonequilibrium Statistical Mechanics},
Solid-State Sciences {\bf 31} (Springer, Berlin, 1991).

\bibitem{karrasch2012}
C. Karrasch, J. H. Bardarson, and J. E. Moore,
Phys. Rev. Lett. {\bf 108}, 227206 (2012).

\bibitem{prosen2011}
T. Prosen,
Phys. Rev. Lett. {\bf 106}, 217206 (2011).

\bibitem{prosen2013}
T. Prosen and E. Ilievski,
Phys. Rev. Lett. {\bf 111}, 057203 (2013).

\bibitem{steinigeweg2014}
R. Steinigeweg, J. Gemmer, and W. Brenig,
Phys. Rev. Lett. \textbf{112}, 120601 (2014).

\bibitem{steinigeweg2015}
R. Steinigeweg, J. Gemmer, and W. Brenig,
Phys. Rev. B \textbf{91}, 104404 (2015).

\bibitem{fabricius1998}
K. Fabricius and B. M. McCoy,
Phys. Rev. B {\bf 57}, 8340 (1998).

\bibitem{steinigeweg2017_1}
R. Steinigeweg, F. Jin, D. Schmidtke, H. de Raedt, K. Michielsen, and J. Gemmer,
Phys. Rev. B \textbf{95}, 035155 (2017).

\bibitem{steinigeweg2017_2}
R. Steinigeweg, F. Jin, H. De Raedt, K. Michielsen, and J. Gemmer,
Phys. Rev. E \textbf{96}, 020105(R) (2017).

\bibitem{karrasch2017}
C. Karrasch, T. Prosen, and F. Heidrich-Meisner,
Phys. Rev. B {\bf 95}, 060406(R) (2017).

\bibitem{karrasch2014}
C. Karrasch, J. E. Moore, and F. Heidrich-Meisner,
Phys. Rev. B {\bf 89}, 075139 (2014).

\bibitem{force} Not for every observable ${\cal O}_l$ there is a force
\cite{luttinger1964, zwanzig1965, gemmer2006}.

\bibitem{luttinger1964}
J. M. Luttinger,
Phys. Rev. {\bf 135}, A1505 (1964).

\bibitem{zwanzig1965}
R. Zwanzig,
Annu. Rev. Phys. Chem. {\bf 16}, 67 (1965).

\bibitem{gemmer2006}
J. Gemmer, R. Steinigeweg, and M. Michel,
Phys. Rev. B {\bf 73}, 104302 (2006).

\bibitem{Essler2014}
F. H. L. Essler, S. Kehrein, S. R. Manmana, and N. J. Robinson, 
Phys. Rev. B {\bf 89}, 165104 (2014).

\bibitem{bartsch2017}
C. Bartsch and J. Gemmer,
EPL (Europhys. Lett.) \textbf{118}, 10006 (2017).

\bibitem{peterson1967}
R. L. Peterson,
Rev. Mod. Phys. {\bf 39}, 69 (1967).

\bibitem{reimann2018}
P. Reimann,
Phys. Rev. E {\bf 97}, 062129 (2018).

\bibitem{Endo2018}
H. Endo, C. Hotta, and A. Shimizu, 
Phys. Rev. Lett. {\bf 121}, 220601 (2018). 

\bibitem{herbrych2016}
J. Herbrych and X. Zotos,
Phys. Rev. B {\bf 93}, 134412 (2016).

\bibitem{weisse2006}
A. Wei\ss{}e, G. Wellein, A. Alvermann, and H. Fehske,
Rev. Mod. Phys. {\bf 78}, 275 (2006).

\bibitem{Karrasch2013}
C. Karrasch, R. Ilan, and J. E. Moore, 
Phys. Rev. B {\bf 88}, 195129 (2013).

\bibitem{Prufer2018}
M. Pr\"ufer, P. Kunkel, H. Strobel, S. Lannig, D. Linnemann, 
C-M. Schmied, J. Berges, T. Gasenzer, and M. K. Oberthaler, 
Nature {\bf 563}, 217 (2018). 

\bibitem{Erne2018}
S. Erne, R. B\"ucker, T. Gasenzer, J. Berges, and J. Schmiedmayer,
Nature {\bf 563}, 225 (2018). 

\bibitem{Richter2018_2}
J. Richter, J. Herbrych, and R. Steinigeweg, 
Phys. Rev. B {\bf 98}, 134302 (2018).

\bibitem{richter2018}
J. Richter, J. Gemmer, and R. Steinigeweg,
arXiv:1805.11625.


\end{thebibliography}
\end{document}